\documentstyle[twoside, ibvs2, ibvst, epsf]{article}

\begin{document}
\IBVShead{5226}{00 Month 200x}

\IBVStitle{Discovery of the CV ROTSE3~J015118.59--022300.1}

\IBVSauth{Smith, D. A.$^1$; Akerlof, C.$^1$; Ashley, M. C. B.$^2$; Casperson, D.$^3$; Gisler, G.$^3$; Henden, A.$^4$; Marshall, S.$^5$; McGowan, K.$^3$; McKay, T.$^1$; Phillips, M. A.$^2$; Rykoff, E.$^1$; Shectman, S.$^6$; Vestrand, W. T.$^3$; Wozniak, P.$^3$}

\IBVSinst{2477 Randall Laboratory, University of Michigan, 500 E. Univeristy Ave., Ann Arbor, MI, 48109, USA}
\IBVSinst{School of Physics, University of New South Wales, Sydney 2052, Australia}
\IBVSinst{Los Alamos National Laboratories, Los Alamos, NM, 87545, USA}
\IBVSinst{USRA/USNO Flagstaff Station, P.O Box 1149, Flagstaff, AZ, 86001, USA}
\IBVSinst{Lawrence Livermore National Laboratory, 7000 East Avenue, Livermore, CA, 94550, USA}
\IBVSinst{Carnegie Observatories, 813 Santa Barbara Street, Pasadena, CA, 91101, USA}

\IBVStyp{CV}
\IBVSkey{photometry}
\IBVSabs{We report on the discovery of an outburst from a previously unknown high-}
\IBVSabs{latitude CV we designate ROTSE3 J015118.59-022300.1.  The object was first}
\IBVSabs{detected on 2001 Oct 13.291 at m-ROTSE3=14.71+-0.06, after which it faded}
\IBVSabs{by more than 2 magnitudes over 13 d.  Spectral analysis confirms a Galactic}
\IBVSabs{location, despite unusual properties that render classification difficult.}

\begintext

On October 11, 2001, the first ROTSE-III automated telescope began observations
in Los Alamos, NM, USA.  ROTSE-IIIa is an 0.45~m telescope with a 1.85 degree
field-of-view, carrying an unfiltered, thinned CCD.  Although the primary task
of this telescope is to rapidly respond to satellite detections of Gamma-Ray
Bursts, most of the observing time is used to perform automated sky patrols.
Pairs of images are taken for each of $\sim100$ patrol fields twice a night.

Analysis of the first ROTSE-IIIa dark run has uncovered an interesting
transient event, which we identify as a nova and designate
ROTSE3~J015118.59--022300.1.  This object is absent in images taken on
October~11, 2001, to limiting aperture magnitudes of $m_{\rm ROTSE3}\sim17.9$.
It is also absent from skyview images, scanned SERC plates from the USNO~PMM
archive, and the USNO~A2.0 catalog.  On October 13, 2001, however, the object
appears at $m_{\rm ROTSE3}=14.00\pm0.06$ (weather conditions prohibited
observations on Oct~12), after which it fades rapidly, falling by more than
2~magnitudes over the next 13~d (Fig.~\ref{fig:lc}).  Assuming an onset time of
1.0~d before the first detection, the best-fit decay index for the light curve
is $\alpha\sim0.9$ ($m\propto t^{-\alpha}$), although even the early light
curve in Figure~\ref{fig:lc} is clearly inconsistent with a single power law.

On 2001 Nov~20.129 and Dec~14.235 the source was observed at BVR$_{\rm
c}$I$_{\rm c}$ using the USNO, Flagstaff Station 1.0m telescope with a
SITe/Tektronix $1024\times1024$ CCD camera (Henden~2001).  Figure~\ref{fig:uim}
shows the V-band image from the Nov 20 dataset.  The nova was found to be at
$V=20.70\pm0.07$ and $20.90\pm0.05$ in the two observations, respectively,
indicating the system may have returned to its quiescent state.  This
brightness, however, is still well above the POSS-II blue plate limit; there
may still be some residual activity from the disc, and the quiescent spectrum
may be extremely red.  The four-color photometry for these two observations is
reported in Table~1.  The USNO images along with USNO-A2.0 provide the
following coordinates for the object: R.A.=$01\hr51\mm18\fsec60\pm0.01$ and
Decl.$=-2\deg23\arcm0.\arcs42\pm0.09$ (J2000.0).  The ROTSE-IIIa position is
consistent with this location to within its errors (0$\fsec$06 R.A.,
$0.\arcs8$ Decl.).  The ROTSE astrometric accuracy has been calibrated against
the USNO~A2.0 catalog (Smith et al. 2002).

We used the first V-band image to set a photometric zero-point for the
unfiltered ROTSE-III images and estimate the V magnitude for the source.  Using
46~bright stars ($m_{\rm ROTSE3}<16$) that show no evidence for variations in
intensity ($\sigma_m < 0.2$~mag) during the 56~ROTSE-III observations, we found
the median offset between the ROTSE aperture magnitudes and the USNO V-band
magnitudes to be $V-m_{\rm ROTSE3}=+0.71$.  The offset for any given object is
of course dependent on its spectral energy distribution, and we show the offset
as a function of color for three different colors in Figure~\ref{fig:col}.
Since the spectrum of the nova during the outburst is unknown, this necessarily
introduces undetermined systematic errors into the magnitude estimates.

Figure~\ref{fig:lc} shows as diamonds the ROTSE V-magnitude estimates for
ROTSE3 J015118.59--022300.1.  Also plotted as triangles, late in the light
curve, are the USNO V-band measurements.  Arrows indicate the mean limiting
magnitude for a pair of ROTSE images in which the nova was not detected.  Error
bars on the ROTSE magnitudes include an estimate of the known systematic errors
(as measured through our relative photometry procedure) added in quadrature to
the statistical uncertainty.

At time 02:32:41 (UT) on Nov 11, 2001, a spectrum of the source was recorded in
a 20-minute exposure with the Boller and Chivens Spectrograph on the 6.5~m
Walter Baade Magellan project telescope (Fig.~\ref{fig:spc}).  The spectrum
shows a continuum with broad but relatively weak Balmer emission lines:
H$_\gamma$, H$_\delta$, H$_\epsilon$ and H$_8$.  The lines are about
3000~km~s$^{-1}$ wide and have a ``square'' profile (steep sides and flat top),
characteristic of accretion disc systems.  The intensity of the line emission
is about half that of the continuum.  The radial velocity is less than a few
hundred~km~s$^{-1}$.

We therefore identify ROTSE3~J015118.59--022300.1 as a galactic cataclysmic
variable.  While the high galactic latitude ($b=-40.74^\circ$) of this object
is unusual, it is not unprecedented (Downes, Webbink \& Shara 1997).  If we
classify this event as a fast nova, the scaling relations in Duerbeck (1981)
would predict a peak absolute magnitude around V$\sim-8.5$.  An extrapolation
of the decaying light curve predicts a peak apparent maximum of V$\sim14.5$ at
one day before the first detection, which implies a distance modulus (DM) of
$\sim23$, or 420~kpc.  If our conversion to V-band overestimates V, the
resulting distance could be as low as 320~kpc.  At this latitude, extinction
cannot explain this unreasonably large distance.  If it is a fast nova, it is
an unusually dim one.  With the large increase in brightness, it is unlikely
that the source is a dwarf nova (Osaki~1996).  Also, its absence in the plates
scanned by the USNO~PMM machine (12~epochs from 1953-1997) requires a low duty
cycle.  If this is a dwarf nova, it may be akin to WZ~Sge.  It may be a
recurrent nova: a diverse class known to recur on timescales of decades
(Cordova~1994).  Further observations are necessary to reliably classify this
system.

ROTSE is supported by NASA under SR\&T grant NAG5-5101, the NSF under grant AST
99-70818 and fellowship 00-136, the UofM, and the Michigan Space Grant
Consortium.  Work by LANL is supported by the DoE under contract W-7405-ENG-36.

\vspace{-0.5cm}

\references

Cordova, F. 1994, in {\it X-ray Binaries}, eds. Lewin, W., van Paradijs, J., \&
van den Heuvel, E. (Cambridge Univerisity Press: Cambridge), 331.
 
Downes, R. A., Webbink, R. F., \& Shara, M. M. 1997, {\it PASP}, {\bf 109},
345.\BIBCODE{1997PASP..109..345D}

Duerbeck, H. W. 1981, {\it PASP}, {\bf 93}, 165.\BIBCODE{1981PASP...93..165D}

Henden, A. A. 2001, {\tt ftp://ftp.nofs.navy.mil/pub/outgoing/aah/sequence/rj0151.dat}

Osaki, Y. 1996, {\it PASP}, {\bf 108}, 39.\BIBCODE{1996PASP..108...39O}

Smith, D., {\it et al.} 2002, in {\it Gamma-Ray Burst and Afterglow Astronomy
2001}, eds. G. Ricker \& R. Vanderspeck (AIP:New York), {\it in press}.

\endreferences

\vspace{-0.5cm}

\IBVSefigure{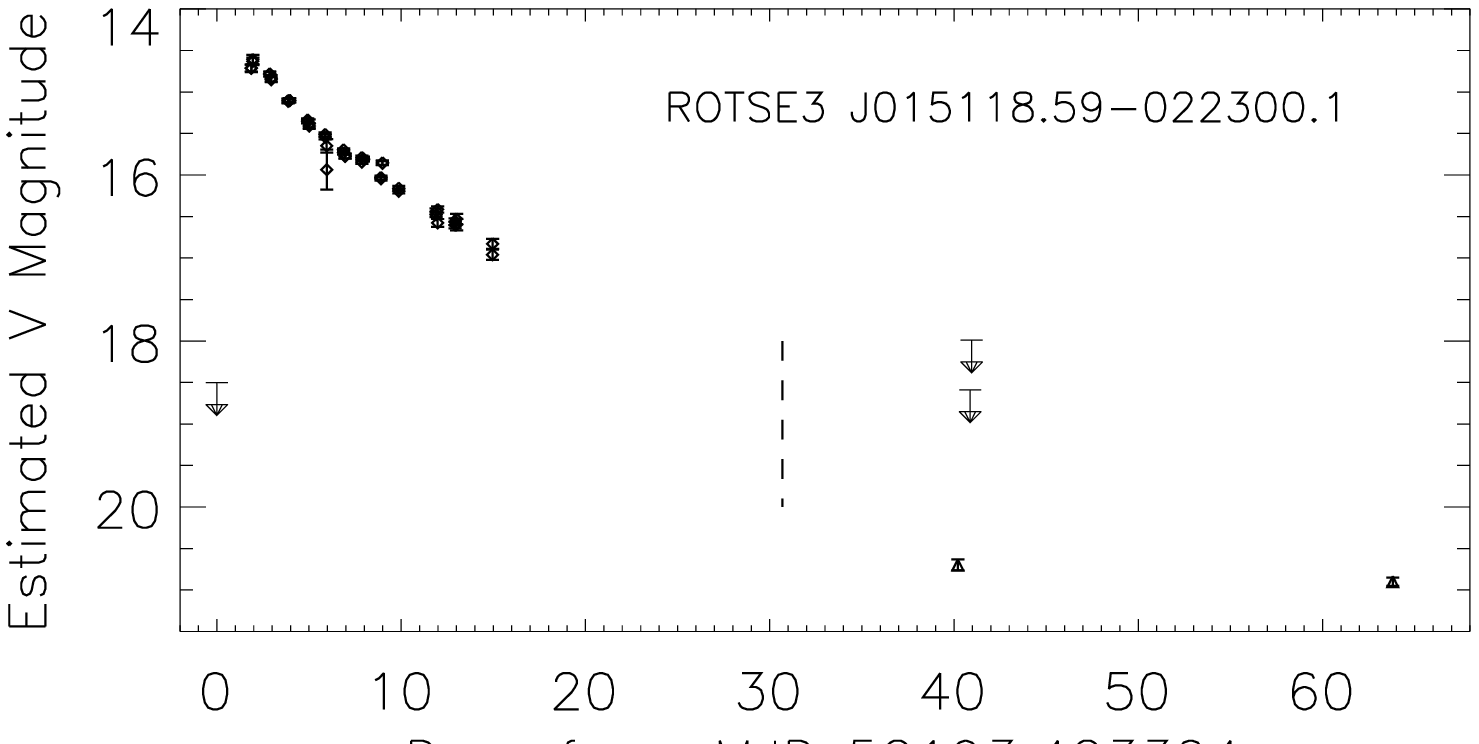}
\IBVSefigure{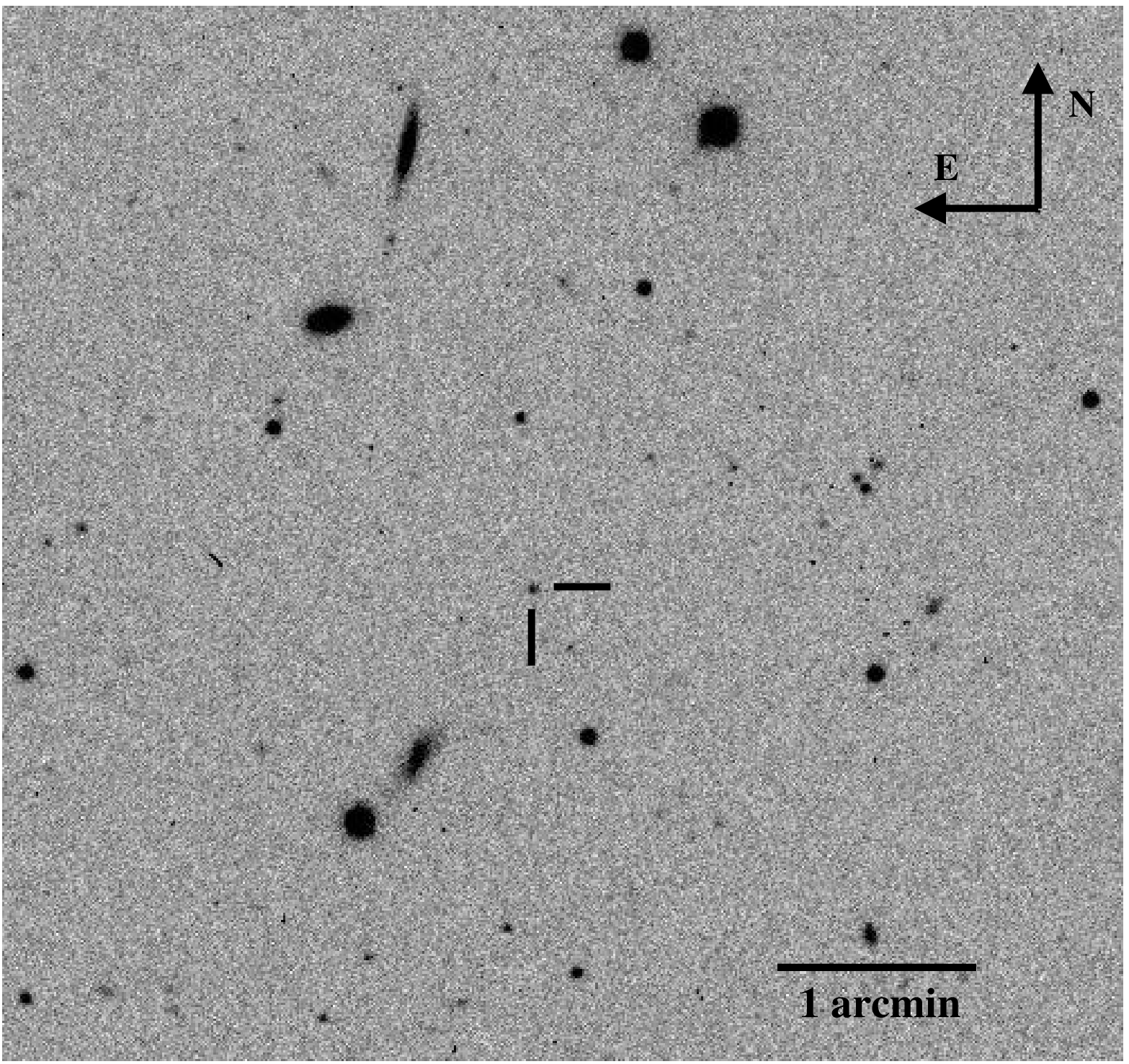}
\IBVSefigure{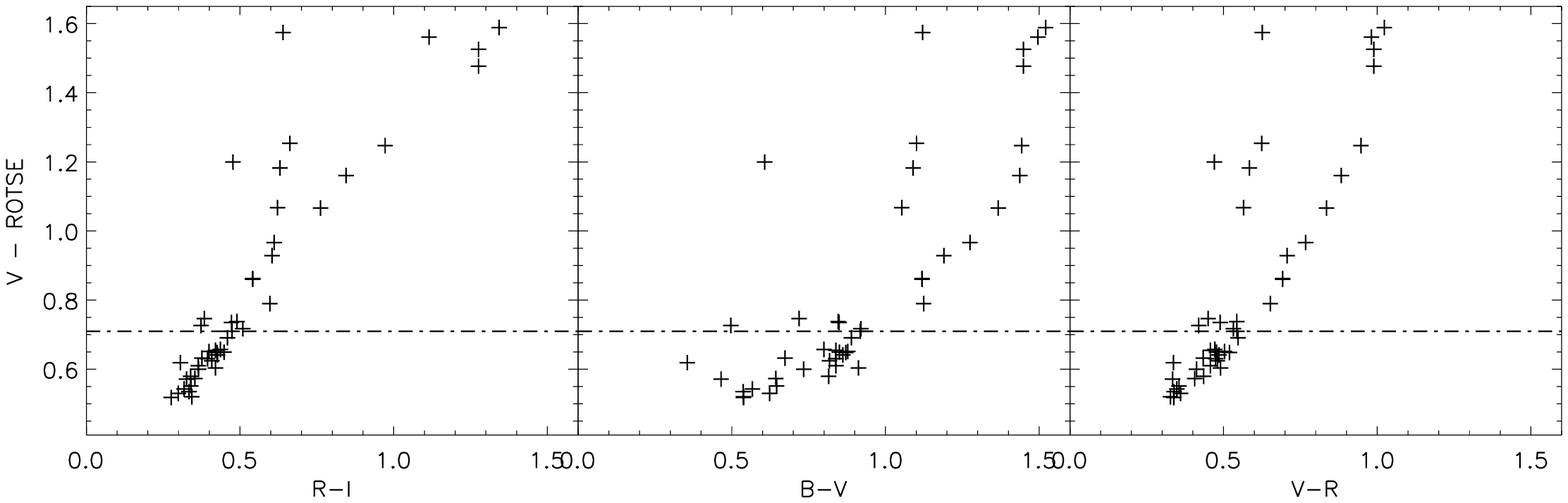}
\IBVSefigure{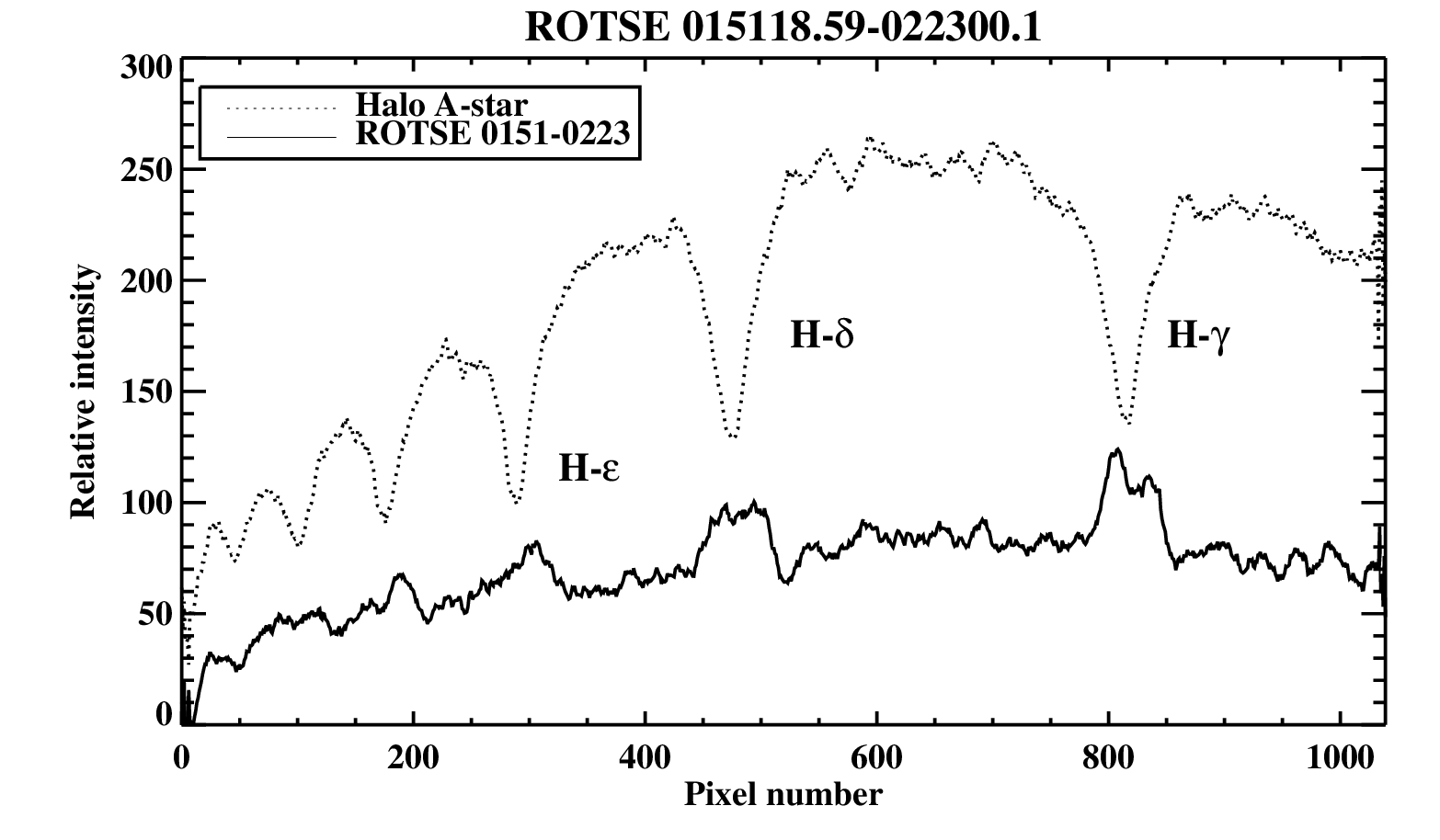}

\IBVSfig{7.5cm}{fig1.eps}{Light curve for a transient nova discovered by
ROTSE-IIIa.  The ROTSE-IIIa unfiltered magnitudes have been corrected by
$+0.71$ to estimate the source's V-band magnitude.  Arrows indicate the mean
limiting magnitudes of pairs of images in which the source was not detected.
V-band observations with the USNO Flagstaff Station 1.0m telescope are
indicated by the triangles at 40 and 64 days. The vertical dashed line
indicates when a spectrum of the source was taken with Magellan.\label{fig:lc}}

\IBVSfig{8.5cm}{fig2.eps}{Image of the region around
ROTSE3~J015118.59--022300.1, from the USNO Flagstaff Station 1.0m
telescope at 2001 Nov~20.129~(UTC).  The nova, at $V=20.70\pm0.07$, is
indicated by crosshairs.\label{fig:uim}}

\IBVSfig{5cm}{fig3.eps}{Offset between ROTSE magnitude and V magnitude
versus the BVRI colors for 46 template stars of constant intensity.  The broken
line indicates the median value of this offset.\label{fig:col}}

\IBVSfig{8.2cm}{fig4.eps}{Spectrum of ROTSE3~J015118.59--022300.1, as compared
with a nearby halo A-star, from a 20-minute exposure with the Boller and
Chivens Spectrograph on the 6.5~m Walter Baade Magellan project telescope at
2001 Nov~11.1060~(UTC).\label{fig:spc}}

\centerline{Table 1. The USNO Four-Color Intensity Measurements for ROTSE3~J015118.59--022300.1}
\vskip 3mm
\begin{center}
\begin{tabular}{ccccc}
\hline
UTD   &     B    &     V   &    R    &   I   \\
\hline
  011121.129  & $20.74\pm0.05$ & $20.70\pm0.07$ & $20.25\pm0.06$ & $19.80\pm0.08$ \\
  011214.235  & $20.91\pm0.03$ & $20.90\pm0.05$ & $20.76\pm0.07$ & $20.29\pm0.10$ \\
\hline
\end{tabular}
\end{center}

\end{document}